\documentclass[journal=jacsat,manuscript=article]{achemso}

\usepackage[version=3]{mhchem}
\usepackage{xcolor}

\author{Vadim Neklyudov}
\affiliation{Wolfson Department of Chemical Engineering, Technion - IIT, Haifa 32000, Israel}
\author{Viatcheslav Freger}
\affiliation{Wolfson Department of Chemical Engineering, Technion - IIT, Haifa 32000, Israel}
\alsoaffiliation{Russel Berrie Nanotechnology Institute, Technion - IIT, Haifa 32000, Israel}
\alsoaffiliation{Grand Technion Energy Program, Technion - IIT, Haifa 32000, Israel}
\email{vfreger@technion.ac.il}
\phone{+972 (0) 4 829 2933}

\title[An \textsf{achemso} demo]
  {Water and Ion Transfer to Narrow Carbon Nanotubes: Roles of Exterior and Interior}

\begin{document}

\begin{tocentry}
\center{\includegraphics[width=0.55\linewidth]{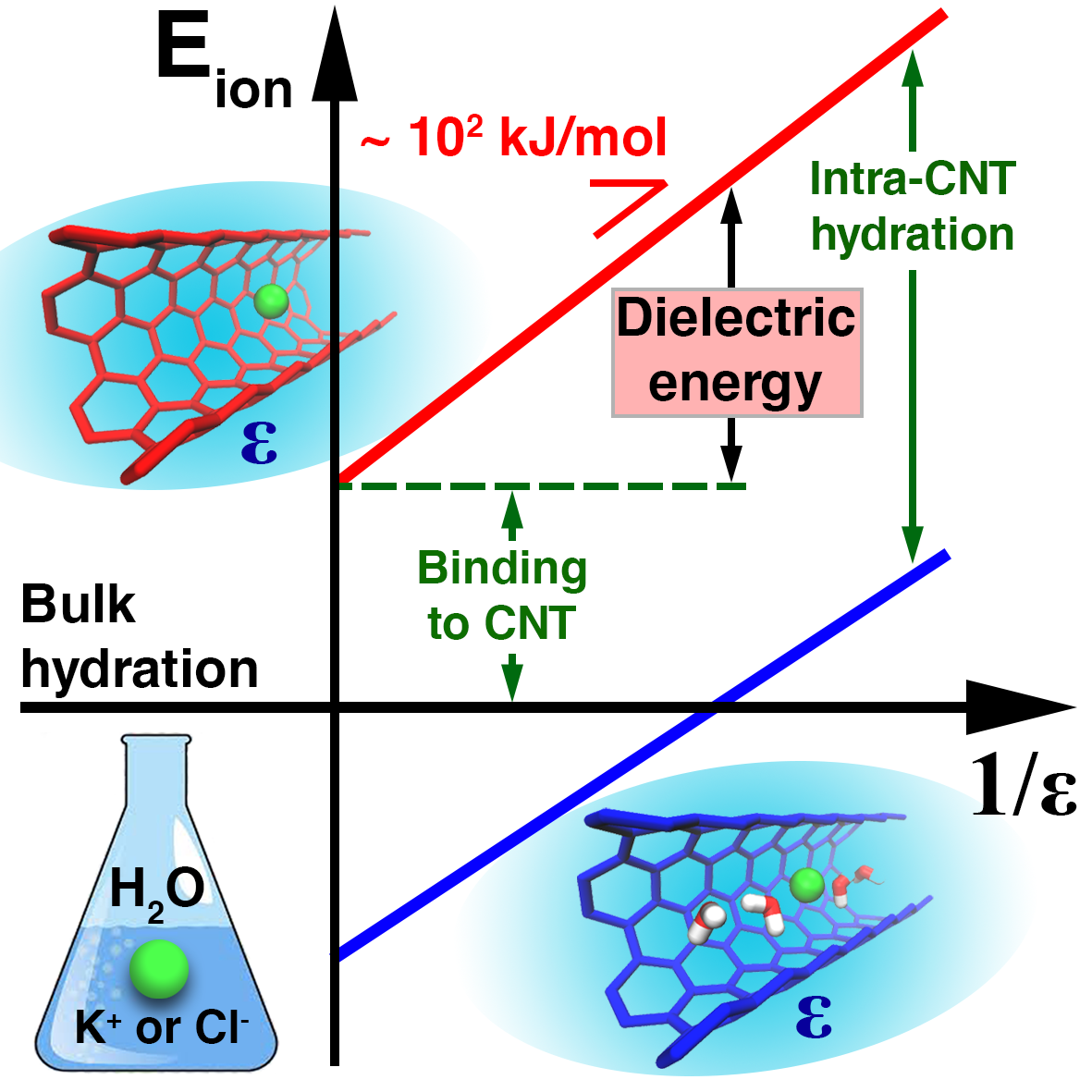}}
\end{tocentry}

\begin{abstract}
Narrow carbon nanotubes (CNTs) desalinate water, mimicking water channels of biological membranes, yet the physics behind selectivity, especially, the effect of the membrane embedding CNTs on water and ion transfer, is still unclear. Here, we report \textit{ab initio} analysis of the energies involved in transfer of water and K\(^{+}\) and Cl\(^{-}\) ions from solution to empty and water-filled 0.68 nm CNTs, for different dielectric constants \(\epsilon\) of the surrounding matrix. The transfer energies computed for 1 \(\leq \epsilon < \infty\) permit a transparent breakdown of the transfer energy to three main contributions: binding to CNT, intra-CNT hydration, and dielectric polarization of the matrix. The latter scales inversely with \(\epsilon\) and is of the order
\(10^{2}\)/\(\epsilon\) kJ/mol for both ions, which may change ion transfer from favorable to unfavorable, depending on ion, \(\epsilon\), and CNT diameter. This may have broad implications for designing and tuning selectivity of nanochannel-based devices. \end{abstract}

Living cells desalinate water using specialized protein channels (aquaporins) embedded in cell membranes.\cite{agre2002aquaporin} Inner channels of carbon nanotubes (CNTs) show much similarity with aquaporins, making CNT porins (CNTPs) attractive for next-generation water purification devices, as well as other bio- and nanotechnology applications.\cite{dittrich2006lab,ghost2003carbon, prezdo2011} Extensive experimental \cite{tunuguntla2017enhanced,holt2006fast,hinds2004gavalas,secchi2016scaling,majumber2005enhanced} and theoretical \cite{corry2008designing, hummer2001nature,striolo2006mechanism,aluru2006,prezdo2012} studies shed light on intriguing mechanism of water flow via CNTs. In CNTs wider than about a nanometer, water displays essentially a bulk-like behavior.\cite{holt2006fast} However, similar to graphene nanoslits \cite{keerthi2018ballistic}, atomically smooth inner walls of narrow sub-nanometer CNTs minimize de Brogli scattering and allow an exceptionally fast single-file water transport at rates greatly exceeding hydrodynamic predictions.\cite{kalra2003osmotic,holt2006fast}

Narrow CNTs also exclude ions displaying selectivity on par with today's desalination membranes \cite{freger2018selectivity,tunuguntla2017enhanced}. Charge repulsion by ionized groups at CNT rims or ions adsorbed on inner walls was suggested as a possible ion exclusion mechanism .\cite{secchi2016scaling,fornasiero2008ion,holt2006fast,tunuguntla2017enhanced,corry2008designing,grosjean2016chemisorption} Indeed, continuum models of charged nanochannels reasonably describe ion transport in wide CNTs that display a moderate selectivity.\cite{secchi2016scaling,fornasiero2008ion, biesheuvel2016analysis} Yet, high selectivity of sub-nanometer CNTPs apparently involves distinctly different physics and, in particular, dielectric exclusion, originating from polarization of the medium surrounding the ion.\cite{freger2018selectivity} This energy reduces compensation for ion dehydration upon transfer from bulk solution to CNT interior and raises transfer energy.\cite{Corry2009,razi2020,li2020strong} Nevertheless, despite recent progress,\cite{Noyacsnano2020,Noyscienceadv2020} understanding all contributions to the energy of ion transfer to narrow CNTs, including dielectric effects, is still incomplete.   

A noteworthy point is that polarizing fields may readily cross the CNT walls and extend to CNT exterior. For instance, molecular dynamics (MD) simulations showed that a charge placed next to a CNT could stop water flow,\cite{li2007electrostatic} and even weaker dipolar interactions may similarly modulate it.\cite{gong2008enhancement} Simulations also find that the effect is reciprocal and water moving in the channel may drag along surrounding molecules.\cite{wang2006coulombic} These results strongly suggest that the medium around a CNT may affect its selectivity. Since, like aquaporins in cell membranes, CNTPs need to be embedded in an insulating host membrane or matrix, understanding the effect of such matrix is crucial both for understanding the selectivity mechanism and for designing optimal CNTP-based systems. Here we investigate this aspect, which has not been examined systematically, focusing on the relation between the dielectric constant \(\epsilon\) of the matrix hosting CNTPs  and the transfer energies for water molecules and K\(^{+}\) and Cl\(^{-}\) ions, for which data on cation-anion selectivity in CNTPs are available.\cite{tunuguntla2017enhanced} 
As the parametrization used in classical MD simulations might not capture intricacies of ion-water-CNT interactions,\cite{kalra2003osmotic,liu2016simulated} we resort to computing the appropriate energies \textit{ab initio}. Since our main interest is in single file arrangement, not always guaranteed in CNTs even as narrow as 0.8 nm,\cite{Noyacsnano2020} we select 0.68 nm CNT, about the narrowest ones to allow passage of water.\cite{grosjean2016chemisorption}
\begin{figure}[t]
\center{\includegraphics[width=0.7\linewidth]{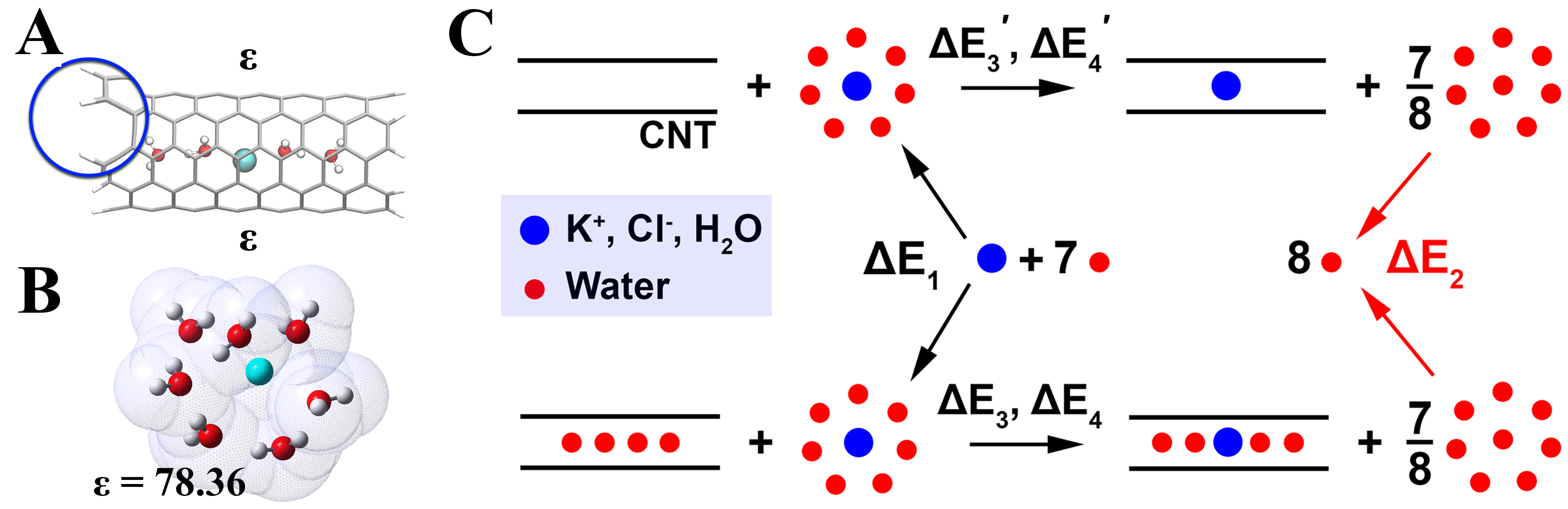}}
\caption{\textbf{(A)} The  simulated cell cut out of optimized infinite CNT and terminated with hydrogen atoms. \textbf{(B)} The  finite cluster of seven water molecules around the species of interest embedded in a dielectric continuum with \(\epsilon\) = 78.36 simulating bulk hydration. The solvent-accessible surface is also shown. Gray, red, white and cyan are carbon, oxygen, hydrogen atoms and the species of interest (H\(_{2}\)O molecule, K\(^{+}\) or Cl\(^{-}\) ions), respectively. \textbf{(C)} Thermodynamic cycles for computing transfer energies for different species between gas phase, bulk water, an empty CNT, and a water file within a CNT.}
\label{fig1}
\end{figure}

The analyzed cell was a (5,5) metallic CNT of length 1.73 nm and diameter 0.68 nm shown in Fig. \ref{fig1}A. Its initial structure without or with water molecules and ions inside was first generated in SIESTA \cite{soler2002ia} at the PBE/DZP level under periodic boundary conditions, as part an infinite CNT(5,5). The structure was optimized at 0 K, equilibrated using \textit{ab initio} MD at 298.15 K, five different snapshots were then re-optimized and the structure with the lowest energy was selected. It was also verified that five species-long single file does not exert excessive internal stress, distorting the cylindrical symmetry of the optimized CNT cell. For subsequent calculations, a 1.73 nm long fragment was cut out and dangling bonds at the rims were terminated with hydrogens  (Fig. \ref{fig1}A). To facilitate computations, we employed B3LYP approximation  with 6-31G(d) basis implemented in GAUSSIAN 09 Rev. B.01.\cite{lee1988development,Gaussian09} The environment around CNT was simulated using the polarizable continuum model IEFPCM.\cite{tomasi1999ief} Bulk hydration energies were computed by optimizing the cluster of the species of interest surrounded by seven water molecules and embedded in a continuum with \(\epsilon\) = 78.36 (Fig. \ref{fig1}B). Previous studies and our computations (Figure S4 in Supporting Information) showed that this cluster size reasonably represents bulk coordination for K$^+$ and Cl$^-$ and well reproduces their hydration, \cite{mahler2012,mancinelli2007,rao2008,zhu2016,robertson2003,bankura2015,bajaj2019} and may suit water as well  .\cite{mejias2000calculation,li2011wat} 
The thermodynamic cycles used to derive transfer energies are schematically shown in Fig \ref{fig1}C. We first benchmarked energy computations versus experimental enthalpies of ion hydration \(\Delta E_{1}\) (transfer from vacuum to bulk water) and water vaporization \(\Delta E_{2}\), computed as follows \cite{mejias2000calculation,uudsemaa2004calculation}
\begin{equation}
\Delta E_1 = E[X(H_2O)_7] - E[X] - 7 (E[(H_2O)_8]/8),
\label{eq1}
\end{equation}
\begin{equation}
\Delta E_2 = E[H_2O] - E[(H_2O)_8]/8,
\label{eq2}
\end{equation}
where X = K\(^{+}\) or Cl\(^{-}\), single species (X or H\(_{2}\)O) are in vacuum and corresponding clusters are embedded in a water-like continuum (Fig. \ref{fig1}B). Note that the above relations approximate the respective excess free energies with enthalpies. This ignores excess entropies, which may be a significant factor in filling 0.8-1 nm CNTs with water,\cite{pascal2011,waghe2012} yet, in the present case, especially, for ion transfer, such entropic contributions would be fairly small compared to enthalpies, including dielectric contribution. The computed \(\Delta E_{1}\) = -318 kJ/mol for K\(^{+}\) and \(\Delta E_{2}\) = 58 kJ/mol were reasonably close to respective experimental values, -334 and 44 kJ/mol.\cite{marcus2015ions,codata} For Cl\(^{-}\) the computed  \(\Delta E_{1}\) = -300 kJ/mol underestimated the experimental value -367 kJ/mol,\cite{ben1984solvation} however, similarly underestimates for Cl\(^{-}\) were found in other studies \cite{li2020strong,mejias2000calculation}. Apparently, the discrepancies reflect the level of theory, especially, unaccounted for dispersion correction. Indeed, we found that extended basis sets produced only a marginal difference, $<$ 5 kJ/mol, yet agreement much improved for the WB97XD functional \cite{chai2008} with Grimme's dispersion correction.\cite{grimme2010} Nevertheless, such improvements insignificantly affected the dielectric energy, as expected from its long-range nature, therefore faster computations were preferred. (See Figures S4 to S6 in Supporting Information for more details.)  

To differentiate the interactions with CNT, dielectric energy and intra-CNT hydration, we computed the transfer energies for two scenarios: (I) insertion of the species into a water file within CNTP and (II) transfer of a single species to an empty CNTP. In scenario I, we presumed that proximity to a water file terminus (“surface energy” of the file, ca. 17 kJ/mol, see Supporing Information) substantially affects only two terminal molecules at each terminus. We then approximated solvation in a long file by that in the middle position of a 5-member file. For both scenarios the transfer energies, \(\Delta E_{3}\) for ions and \(\Delta E_{4}\) for water, were then computed as follows (Fig. \ref{fig1}C) 
\begin{equation}
\Delta E_3 = E[X(H_2O)_4\: in\: CNT] + 7 (E[(H_2O)_8]/8)-E[(H_2O)_4\: in \: CNT] - E[X(H_2O)_7],
\label{eq3}
\end{equation}
\begin{equation}
\Delta E_4 = E[(H_2O)_5\: in\: CNT] - E[(H_2O)_8]/8-E[(H_2O)_4\: in \: CNT].
\label{eq4}
\end{equation}
In all initial and final states, CNT was embedded in a dielectric continuum with a corresponding \(\epsilon\). Scenario I was addressed by considering the four terminal water molecules as real molecules (energies \(\Delta E_{3}\) and \(\Delta E_{4}\)), whereas in scenario II (energies \(\Delta E_{3}^{'}\) and \(\Delta E_{4}^{'}\)) they were non-interacting dummy molecules that excluded the external continuum and kept \(\epsilon = 1\) inside CNT.

\begin{figure}[t]
\center{\includegraphics[width=1\linewidth]{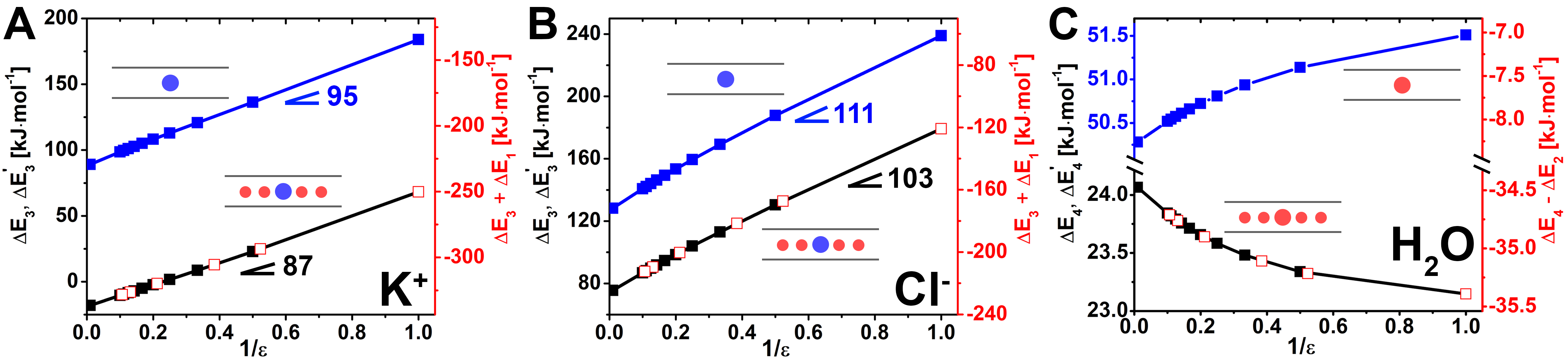}}
\caption{Variation of the transfer energies for K\(^{+}\) \textbf{(A)} and Cl\(^{-}\) \textbf{(B)} ions and a water molecule \textbf{(C)} with \(1/{\epsilon}\). The left and right vertical axes correspond to transfer from bulk water and from vacuum, respectively. Blue lines represent transfer of the respective ion (\(\Delta E_{3}^{'}\)) or water molecule (\(\Delta E_{4}^{'}\)) into an empty CNT. Black lines represent insertion of the respective ion (\(\Delta E_{3}\)) or water molecule (\(\Delta E_{4}\)) in water file within a CNT. Indicated are the slopes of the lines. The filled symbols are results computed for CNT embedded in a dielectric continuum. The empty red symbols represent CNT embedded (in the order of increasing \(\epsilon\)) in vacuum and effective media representing common solvents (heptane, hexanoic acid, chloroform, iodoethane, and 2-bromopropane).}
\label{fig2}
\end{figure}
Figures \ref{fig2}A and B display the key result, the ion transfer energies \(\Delta E_{3}\) (water-filled channel, black lines) and \(\Delta E_{3}^{'}\)  (empty tube, blue lines) plotted versus \(1/{\epsilon}\). Here, the zero energy corresponds to ion bulk dehydration, \(-\Delta E_{1}^{'}\), different for each species.  Empty symbols are results for the dielectric continuum around CNT replaced with effective media representing various common solvents and including various \textit{ad hoc} corrections for solvent-specific short-range non-dispersive interactions (hydrogen bond acidity and basicity, aromaticity, electronegative halogenicity). The fact that \textit{ad hoc} media closely match the trend of dielectric continua indicates that \(\epsilon\) of the exterior is the main factor controlling its interaction with the ion. This could be expected, as CNT walls efficiently eliminate such short-range interactions.  For both ions, \(\Delta E_{3}\) and \(\Delta E_{3}^{'}\)  show similar linear dependence with slopes of the order \(10^{2}\) kJ/mol, only slightly different for each ion and scenario. The linear dependence on \(1/{\epsilon}\) is suggested by the Born solvation energy of an ion a dielectric continuum
\begin{equation}
E_{Born} = \frac{e^2}{8\pi \epsilon_0 \epsilon R},
\label{eq5}
\end{equation}
where \(e\) is the electron charge, \(\epsilon_{0}\) the permittivity of vacuum, and \(R\) is the effective radius of an equivalent ideally polarizable spherical cavity containing the ion, identified as the ion radius in the simple theory. 

In contrast to previous computations of transfer energies for a specific surrounding (usually, vacuum \(\epsilon = 1\) or a lipid membrane \(\epsilon \approx 2\)), the entire dependence of \(\Delta E_{3}\) and \(\Delta E_{3}^{'}\) on \(1/{\epsilon}\) obtained here permits a transparent breakdown of the transfer energy to distinct contributions. Specifically, it differentiates the dielectric energy of the exterior from intra-CNT interactions. Indeed, the exterior polarization vanishes for \(1/{\epsilon}\)  = 0 (\(\epsilon = \infty\)) therefore the intercept of \(\Delta E_{3}^{'}\) (the leftmost point on the blue line) reflects only ion binding to bare CNT relative its dehydration energy, \(-\Delta E_{1}\). It is about 89 kJ/mol for K\(^{+}\) and 128 kJ/mol for Cl\(^{-}\) in present computations and the increase of \(\Delta E_{3}^{'}\) above this value for finite \(\epsilon\), varying about linearly with \(1/{\epsilon}\) (eq. \ref{eq5}), represents the dielectric polarization of the exterior. This contribution substantially increases the transfer energy, reaching its maximum, about +184 kJ/mol for K\(^{+}\) and +239 kJ/mol for Cl\(^{-}\), at \(\epsilon\) = 1 (vacuum). Finally, the vertical separation of the black and blue lines, \(\Delta E_{3} - \Delta E_{3}^{'}\), is the energy of the intra-CNT ion hydration. Although it is only a fraction of the bulk hydration, it is commensurate with the other contributions and offsets substantially the dehydration penalty. 

To appreciate the relative role of different contributions, consider \(\epsilon\) = 2, typical of lipid membranes.\cite{dilger1979dielectric} The total transfer energy for K\(^{+}\) ion, 23 kJ/mole, is made up of 89 (ion binding to CNT relative to bulk hydration), -113 (intra-CNT hydration), and +47 kJ/mol (dielectric energy). For Cl\(^{-}\), the respective values sum up as +130 = 128 – 58 + 60 kJ/mol. Although the absolute numbers may change with the level of theory, it is seen that the dielectric energy for K\(^{+}\) is commensurate with the other contributions, therefore potassium transfer to CNT may be both favorable and unfavorable, depending on \(\epsilon\). The insertion of Cl\(^{-}\) in a water-filled CNTP has an energy barrier at any \(\epsilon\) value, however, increasing dielectric constant can significantly reduce it. 

Note that, based on eq. \ref{eq5}, the linear slopes in Figs. \ref{fig2}A and B may be identified as \(e^2/8\pi \epsilon_{0} R_{ef}\) with some effective value of \(R_{ef}\). They yield \(R_{ef}\) = 0.73 nm for K\(^{+}\) and \(R_{ef}\) = 0.62 nm for Cl\(^{-}\), which significantly exceed and inversely correlate with the respective bare ion radii \(R_{ion}\), 0.141 and 0.180 nm.\cite{marcus1988ionic} The difference between \(R_{ef}\) and \(R_{ion}\) reflects the CNT electron cloud polarization by the ion as well as ion charge delocalization due to ion bonding with CNT. Apparently, the smaller K\(^{+}\) more strongly binds to CNT, resulting in larger \(R_{ef}\), compared to Cl\(^{-}\). The fact that \(R_{ef}\) is intermediate to the CNT radius (0.34 nm) and half-length (0.865 nm) suggests that polarization and charge delocalization extend along the entire CNT, distorting spherical symmetry and reducing intensity of the ion field. The small difference between the slopes of \(\Delta E_{3}\) and \(\Delta E_{3}^{'}\) indicates an additional weak polarization of water within CNT, which slightly increases \(R_{ef}\) to 0.79 and 0.63 nm, respectively. 

\begin{figure*}[t]
\center{\includegraphics[width=1\linewidth]{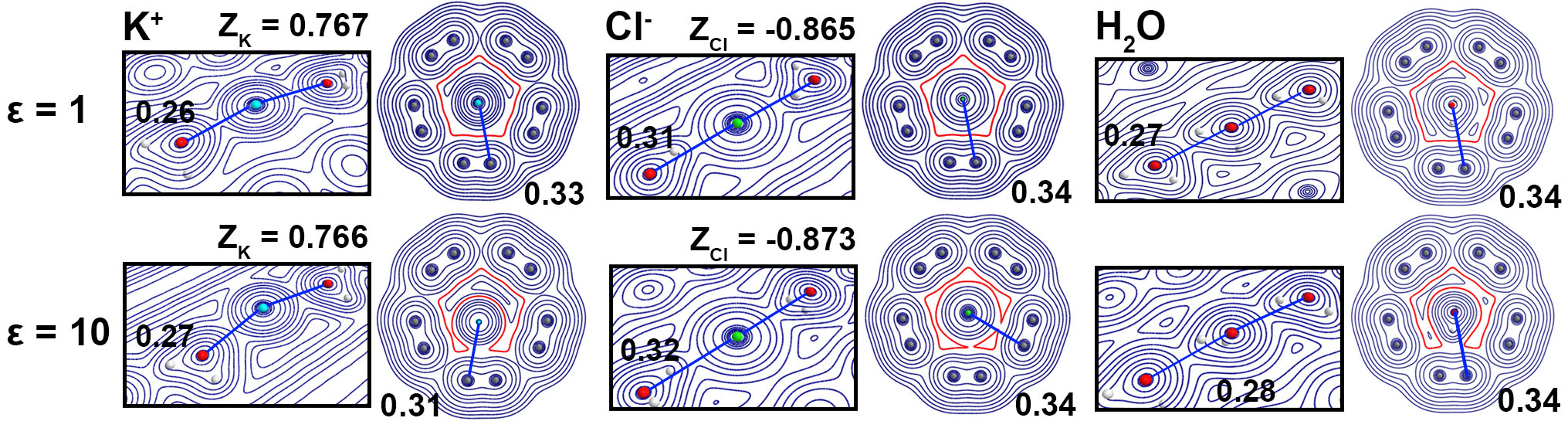}}
\caption{Electron density distribution along and across CNT around hydrated K\(^{+}\), Cl\(^{-}\), and H\(_{2}\)O for \(\epsilon\)=1 (top row) and \(\epsilon\)=10 (bottom row). The numbers within the maps are indicated distances in nanometers. \(Z_{K}\) and \(Z_{Cl}\) are residual ion charges.}
\label{fig3}
\end{figure*}

The above effects may be more explicitly observed in electron density maps in Fig. \ref{fig3} and, specifically, the interatomic distances and the residual ion charge \(Z_{ion}\) evaluated in GAUSSIAN using the natural bond orbital analysis.\cite{glendening2003nbo} K\(^{+}\) shows stronger binding and charge delocalization, manifested in a  smaller residual charge , \(Z_{K} \approx\) +0.77, and shorter K-C distance, compared with \(Z_{Cl} \approx\) -0.87 and Cl-C distance. Notably, the maps for \(\epsilon\) = 1 and \(\epsilon\) = 10 indicate that, at higher \(\epsilon\), K\(^{+}\) bonds more strongly to CNT and more weakly interacts with water (cf. longer ion-water oxygen distance), which is less pronounced for Cl\(^{-}\) and consistent with its smaller \(R_{ef}\).

Compared to ions, transfer energies for H\(_{2}\)O \(\Delta E_{4}\) and \(\Delta E_{4}^{'}\) (Fig. \ref{fig2}C) show a much weaker dependence on \(\epsilon\), as expected of a dipole. While the decreasing trend of \(\Delta E_{4}^{'}\) versus \(1/{\epsilon}\) is reminiscent of ions' \(\Delta E_{3}^{'}\), it reverses for \(\Delta E_{4}\). The difference between \(\Delta E_{4} - \Delta E_{4}^{'}\) reflects the water-water interaction, slightly weakened at higher \(\epsilon\). This is also observed in electron density maps in Fig. \ref{fig3} as the larger distance between water oxygens and weaker bonding at \(\epsilon = 10\), relative to \(\epsilon = 1\). Ultimately, opposite of ions, water transfers to CNTs is somewhat promoted rather than impeded by a lower-dielectric surrounding. 

As explained before, the present choice of 0.68 nm nanotubes, presumably the narrowest permeable,\cite{corry2008designing} was to ensure the single file arrangement, but the experimental data for this diameter are unavailable. Since 0.8 nm CNTPs may preserve or only partly distort the single file, some trends should be qualitatively similar to 0.68 nm channels. Indeed, reversal potential measurements in 0.8 nm CNTPs indicate a K:Cl selectivity ca. 200:1,\cite{tunuguntla2017enhanced} which is fully consistent with the far more favorable transfer energy of K\(^{+}\) vs. Cl\(^{-}\) found here. Similarly, water transfer is more favorable compared to Cl\(^{-}\), the ion that has the larger transfer energy and controls salt transfer, i.e., CNTPs should desalinate water, as observed for 0.8 nm tubes.\cite{tunuguntla2017enhanced} We anticipate that in wider tubes water-water interaction within the file, \(\Delta E_{4}^{'} - \Delta E_{4} \approx\) -28 kJ/mol, should not change significantly, but short-range dipolar interactions of water with CNT (\(\Delta E_{4}\)) may weaken, further facilitating water transfer. 

On the other hand, increased CNT diameter may oppositely affect ion transfer in 0.8 nm nanotubes. Specifically, \(R_{ef}\) should about linearly correlate with the CNT diameter. The 15\% difference between 0.68 and 0.8 nm CNTs may then reduce the dielectric energy, inversely related to \(R_{ef}\), by up to 15 kJ/mol depending on \(\epsilon\). It is more speculative to project to wider channels the ion-CNT interaction, yet, if they are viewed as adsorption on inner walls, the biding energy is expected to scale roughly as the surface area per CNT volume, i.e., about inversely depend on the CNT diameter, similar to dielectric energy. As an illustration, we may project the present data for 0.68 nm to 0.8 nm CNTPs embedded in a lipid bilayer of dielectric constant 2.4 \cite{dopc2015} and estimate transfer energies for water and Cl\(^{-}\), as follows. For water, intra-CNT hydration and bulk dehydration are -28 and 58 kJ/mol, respectively, therefore partial dehydration within CNT costs -28 + 58 = 30 kJ/mol. Binding of a water molecule from vacuum to bare 0.68 nm CNT amounts to \(\Delta E_{4}^{'} - \Delta E_{2} \approx\) = -8 kJ/mol (see the right axis in Figure 2C). With insignificant dielectric energy, the dehydration energy unchanged and binding energy from vacuum reduced by 15\%, water transfer energy to 0.8 nm CNTPs would be about 23 kJ/mol. This well matches experimental transport activation energies (22 kJ/mol) as well as \textit{ab initio} computations (20 kJ/mol).\cite{Noyscienceadv2020}

For Cl\(^{-}\) and \(\epsilon = 2.4\), the computed intra-CNT hydration energy \(\Delta E_{3}^{'} - \Delta E_{3} \approx\) =  -60 kJ/mol vs. bulk dehydration 300 kJ/mol may stay unchanged in 0.8 nm CNTP, while ion binding from vacuum \(\Delta E_{3}^{'} + \Delta E_{1} \approx\) -172 (right axis in Fig. 2B for 1/\(\epsilon\) = 0) and dielectric energy 111/2.4 = 46 kJ/mol are to be reduced by 15\%. Ultimately, this yields the transfer energy 137 kJ/mol. This significantly exceeds the experimental activation barrier for Cl\(^{-}\), 52 kJ/mol\cite{Noyscienceadv2020}. The same study reported this value matched the theoretical values 63 kJ/mol computed using a hybrid DFT method. However, their simulations artificially placed highly polarizable graphene sheets at the CNT exterior, which could produce a situation, equivalent to a large \(\epsilon\) in the present model. This amounts to discarding 46 kJ/mol of dielectric energy, which would take their result close to the present estimate. As an alternative explanation, we speculate that the discrepancy might point to a distortion of single  water file around Cl\(^{-}\). Indeed, our preliminary results indicate that Cl\(^{-}\) in 0.8 nm (6,6) CNT is solvated by three rather than two water molecules (see Figure S7 in Supporting Information), which should decrease transfer energy by several tens kJ/mol. Examination of this point using a higher level of theory and other ions is currently underway, but it is well in line with a similar result reported for small cations in 0.8 CNTP.\cite{Noyacsnano2020}

In summary, the present approach, analysing variation of the transfer energy in the entire range 1 \(\leq \epsilon < \infty\), permits a transparent breakdown of the water and ion transfer energies to several distinct contributions. Along with ion-specific effects that favor uptake of K\(^{+}\) vs. Cl\(^{-}\) ions, the analysis highlights significance of ion interactions with the matrix surrounding the CNT, controlled by its dielectric constant \(\epsilon\). The latter contribution is of the order of a few tens to a hundred kJ/mol for 0.68 nm CNT, which may change ion transfer from favorable to unfavorable in response to decreasing \(\epsilon\). It is expected to remain substantial in wider nanotubes, as long as they preserve the single-file water arrangement, though the effect of weaker binding on transport and water-ion selectivity may be more significant. The present results add to the general physical picture of dielectric exclusion as key selectivity mechanism in desalination membranes and nanochannels.

\begin{acknowledgement}
The financial support by a joint grant 2016627 from the United States-Israel Binational Science Foundation (Israel) and National Science Foundation (USA) is acknowledged. The authors thank Aleksandr Noy and Meni Wanunu for discussions and many valuable suggestions.
\end{acknowledgement}

\begin{suppinfo}
Supporting information contains the procedure for building model structures, details of quantum-chemical calculations in a SIESTA, examination of the effect of water and ion/water clusters size and level of theory on computed hydration enthalpies, free energies, and the dielectric energy, the estimations of the "surface energy" of a water file terminus, and the optimized structure of hydrated chloride ion in CNT(6,6).
\end{suppinfo}

\end{document}


\textbf{Modelled CNT supercell}. The supercell used in optimization and QMD is presented in Figure S1A. The length of the supercell along the CNT is equivalent to seven elementary (one benzene ring-long) (5,5) units. The length of one unit is 0.246 nm therefore the supercell size in the longitudinal (z) direction is about 1.72 nm. The supercell size in x and y dimensions was set to 1.80 nm, large enough to eliminate interaction between neighboring supercells. Figure S1B shows the dependence of energy on the x and y spacing, which plateaus above about 1.20 nm; above this distance, the differences in energy drop below thermal energy (27 meV at 298.15 K).

\begin{figure}[h]
\center{\includegraphics[width=0.9\linewidth]{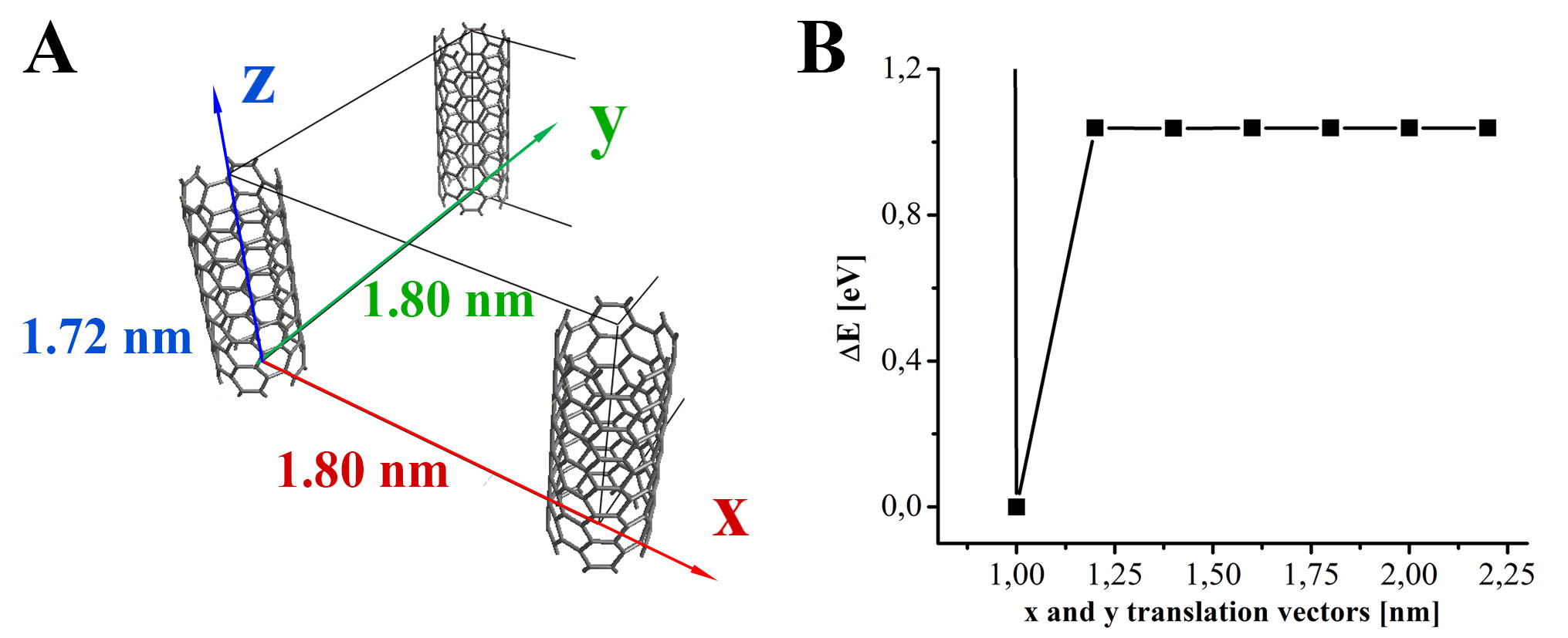}}
\caption{(A) Seven elementary units (1.72 nm) long CNT(5,5) supercell of size 1.72 x 1.80 x 1.80 nm\(^{3}\) of dimensions in infinite, subject to PBC. (B) Relationship between \(\Delta\)E versus lateral (x,y) spacing of CNTs. \(\Delta\)E is defined as the total supercell energy relative to its value at lateral spacing 1.00 nm.}
\label{fig1}
\end{figure}

Different contents of the supercell were simulated by placing an ion (K\(^{+}\) or Cl\(^{-}\)) or a H\(_{2}\)O molecule in the center and surrounding it with four water molecules, two at each side. Such a number of water molecules, on the one hand, reasonably approximates the intra-CNT hydration of the central species in a long file and, on the other hand, leaves a small vacant space within the supercell and thereby avoids stress and distortion of the nanotube. Simulations showed that, for a larger number of molecules within the supercell, there is a transition from a linear single file to a zigzag arrangement and strong distortion of the nanotube cylindrical symmetry, as a result of excessive internal stress. The small residual free volume also allows stress-free adjustment of the occupied volume within the supercell to the different sizes of K\(^{+}\), Cl\(^{-}\) and H\(_{2}\)O (see Figure S2).

\begin{figure}[h]
\center{\includegraphics[width=0.5\linewidth]{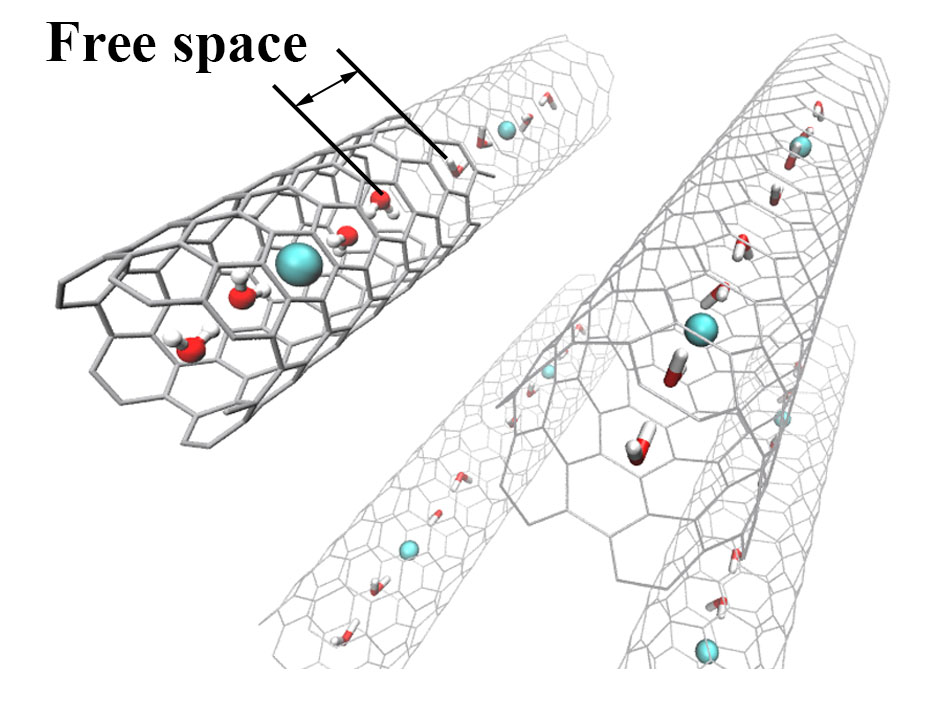}}
\caption{The model structure of an ion or water molecule (cyan-colored) surrounded with four water molecules in the center of a seven (5,5) elementary unit-long supercell.}
\label{fig2}
\end{figure}

\textbf{Selection of simulation parameters in SIESTA.} Before optimization in SIESTA, the values of the parameters Meshcutoff, Splitnorm were selected. The Meshcutoff value 700 Ry was selected for subsequent calculations, which ensures the error of the total energy does not exceed 2 meV. Splitnorm parameters for hydrogen, oxygen, carbon and chlorine atoms were 0.15. For potassium Splitnorm the minimal possible value 0.25 was used. The selected values of parameter were verified by calculating geometry of a few simple benchmark molecules (Figure S3). Deviations from known geometry were below 3\% for bond length and below 1\% for bond angles. The computed distance between neighbored carbon atoms in CNT (5,5) was 0.144 nm, which exactly matches the literature value \cite{jindal2008bond}.

\begin{figure}[h]
\center{\includegraphics[width=1\linewidth]{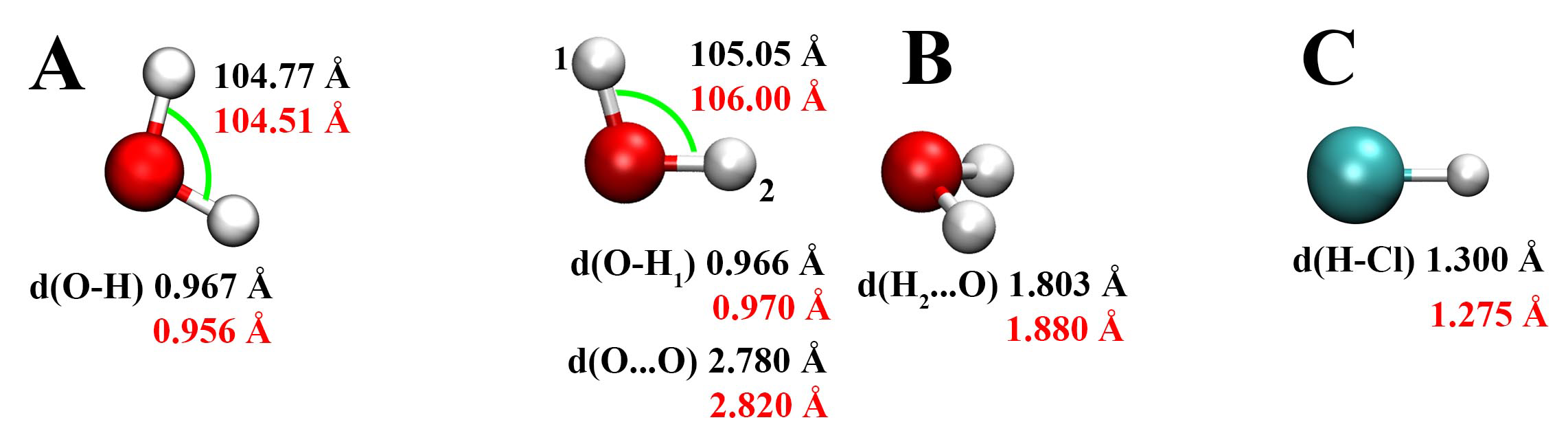}}
\caption{Computed geometry of water molecule (A), water dimer (B) and hydrogen chloride (C). Numbers in red are literature values for water and hydrogen chloride \cite{haynes2014crc} and for water dimer \cite{chaplin2010water}}
\label{fig3}
\end{figure}

\textbf{Optimization in SIESTA and generation of CNT cell for GAUSSIAN.} The finite CNT structure used in this study was generated via optimization and quantum molecular dynamics (QMD) equilibration of an infinite CNT. The optimization using generalized global approximation (GGA) of Perdew-Burke-Erzenhof (PBE) \cite{perdew1996generalized} implemented in SIESTA 3.2 package along with Troullier-Martins norm-conserving pseudopotentials \cite{troullier1991efficient} in the Kleiman-Bylander form \cite{kleinman1982efficacious}. Structural relaxation was based on conjugate gradients with the convergence criterion where forces acting on all atoms do not exceed 0.04  eV \({\AA}^{-1}\) in double-\(\zeta\) polarized (DZP) basis set, translation vectors were also optimized. The non-zero overall charge and charge asymmetry (dipole moment) in periodic boundary conditions (PBC) were compensated using the standard tools and corrections implemented in SIESTA \cite{ordejon1996self}. The QMD equilibration was carried out in NVT assemble using Nose thermostat at 298.15 K temperature. Each analyzed structure (supercell) subject to PBC was first subject to optimization and then to QMD equilibration. Thereafter, five snapshots of QMD-equilibrated structure were re-optimized and the structure with the lowest energy was selected. The selected supercell (part of infinite CNT) was cut out of the optimized infinite CNT and terminated with hydrogen atoms at the rims. The resulting finite cell with an appropriate interior content and surrounded with selected dielectric continua IEFPCM \cite{tomasi1999ief} was used for subsequent simulations in GAUSSIAN at the B3LYP/6-31G(d) level of theory.
\linebreak

\begin{figure}[h]
\center{\includegraphics[width=0.7\linewidth]{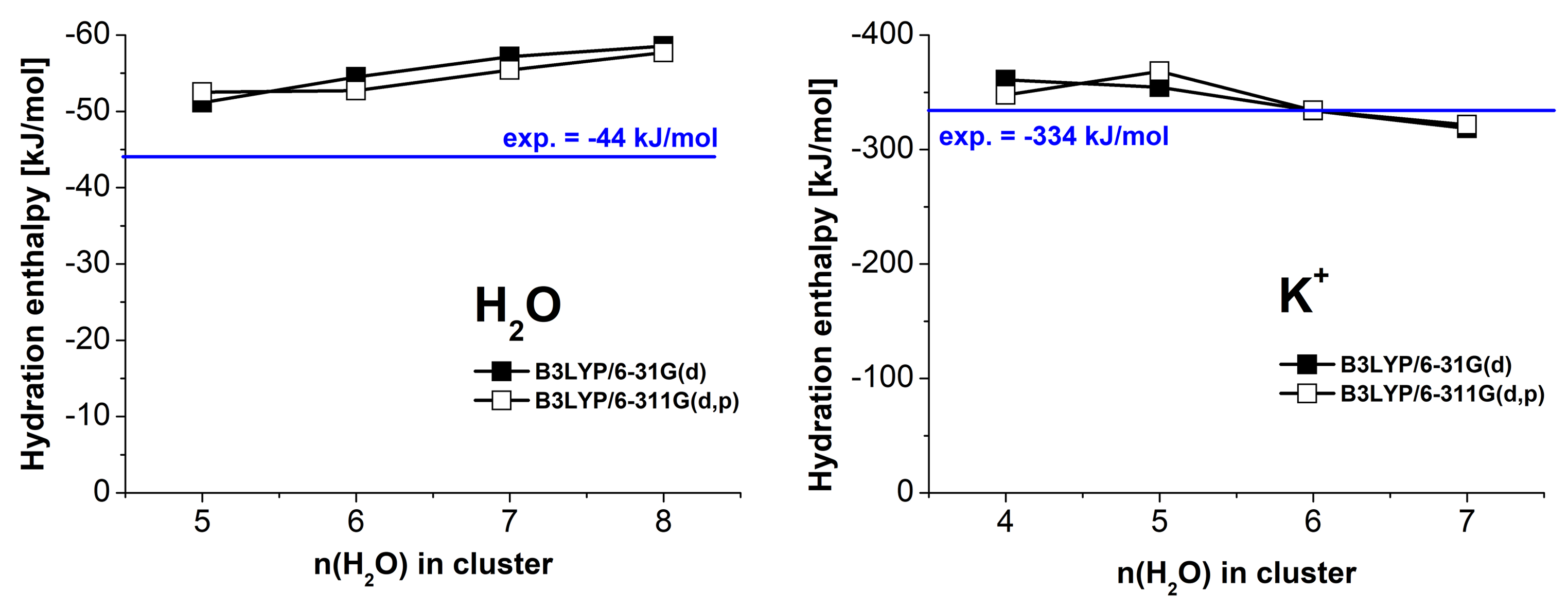}}
\caption{The affect of the number of water molecules in a cluster on the enthalpy of hydration of water and potassium ion.}
\label{fig4}
\end{figure}
\textbf{Effect of cluster size on computed hydration enthalpes.} Figure S4 shows how the number of water molecules surrounding the species of interest in solvated cluster affects the hydration enthaplies of water and potassium ion. In this work, clusters with seven water molecules were selected for calculation of hydration enthalpies of H\(_{2}\)O and K\(^{+}\), respectively. This number yields best agreement with the experimental values of hydration enthalpy for ions since it is close to the bulk coordination number of or K$^+$ (6-7)\cite{mahler2012} and Cl$^-$ (6.1$\pm$1.1)\cite{mancinelli2007}. Using smaller clusters (coordination numbers 4 to 7) does not significantly affect the computed enthalpy,  but addition of more than seven water molecules (coordination numbers 8 and 9) increases deviations from the experiment. Apparently, after filling the first coordination shell, the incompletely filled second coordination shell artificially increases the uncompensated surface energy of the cluster. A similar effect was observed for clusters surrounding Ag$^+$ ion.\cite{martinez1997}

\begin{figure}[h]
\center{\includegraphics[width=1\linewidth]{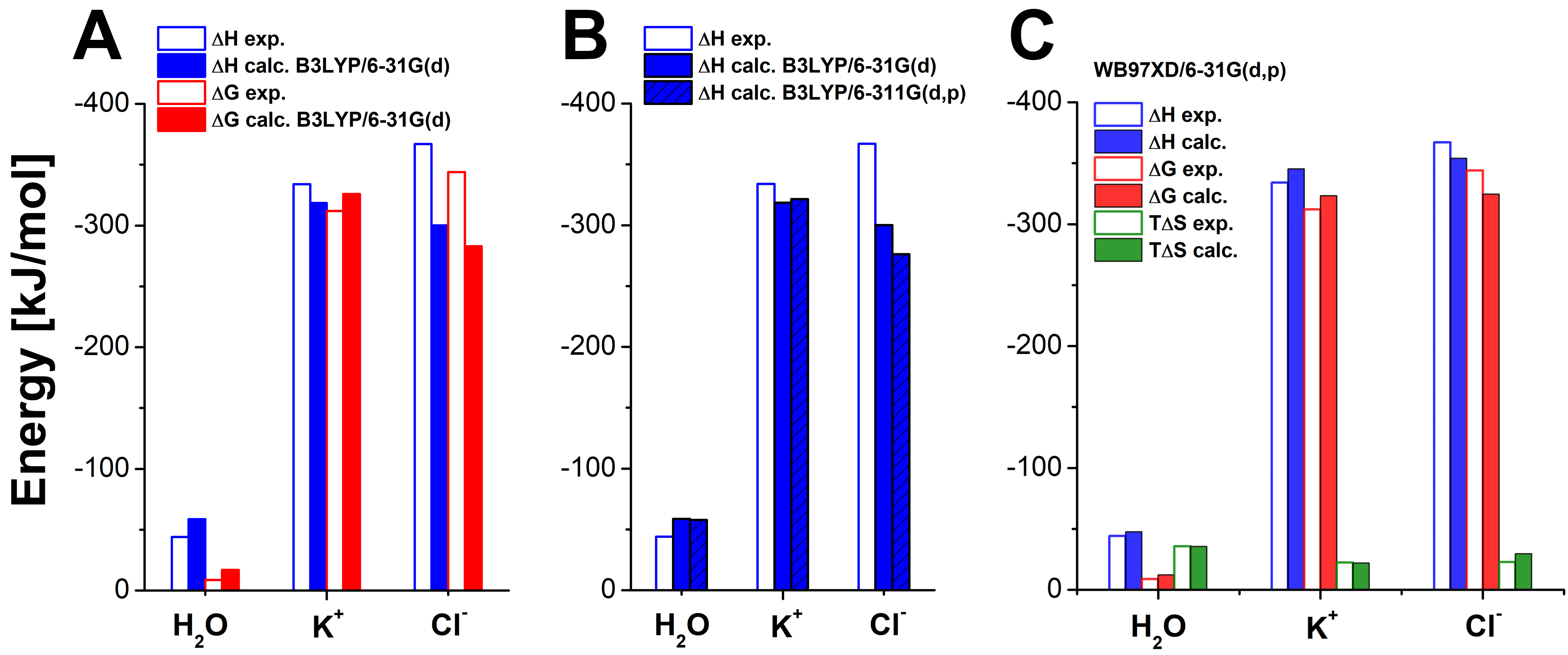}}
\caption{Comparison of the computed and experimental values of bulk hydration enthalpies, excess Gibbs energies and excess enthalpies for H\(_{2}\)O, K\(^{+}\) and Cl\(^{-}\). (\textbf{A}) The experimental hydration enthalpy and excess Gibbs energy compared with computed ones for the basis set B3LYP/6-31G(d) used in this work. (\textbf{B}) The experimental hydration enthalpies compared with those computed using double and triple basis sets B3LYP/6-31G(d) and B3LYP/6-311G(d,p), respectively. (\textbf{C}) The hydration enthalpy, excess Gibbs energy and excess entropy computed in WB97XD/6-31G(d,p) including dispersion correction.}
\label{fig5}
\end{figure}
\textbf{Experimental hydration enthalpies and Gibbs energies versus those computed using different basis sets.} Figure S5 presens comparison of experimental values of bulk hydration enthalpies, excess Gibbs energy and excess entropy for H\(_{2}\)O, K\(^{+}\) and Cl\(^{-}\). Due to relatively small entropic contributions, the  computed entalpies appear to be only marginally different form respective excess Gibbs energies, except for water, however the enthalpy and Gibbs energy for water are much smaller than for ions (Figs. 5SA and C). There is also a reasonable agreement with experimental values, except for Cl\(^{-}\). The agreement is not improved  by using  that the triple basis sets 6-311G(d,p) instead of double set B3LYP/6-31G(d) (Figure 5SB). However, Fig. S5C shows that a good agreement was obtained by using the WB97XD functional \cite{chai2008} in combination with the 6-31G(d,p) basic set, which is a long-range corrected hybrid functional, including the Grimme dispersion correction \cite{grimme2010}. 

\begin{figure}[h]
\center{\includegraphics[width=1\linewidth]{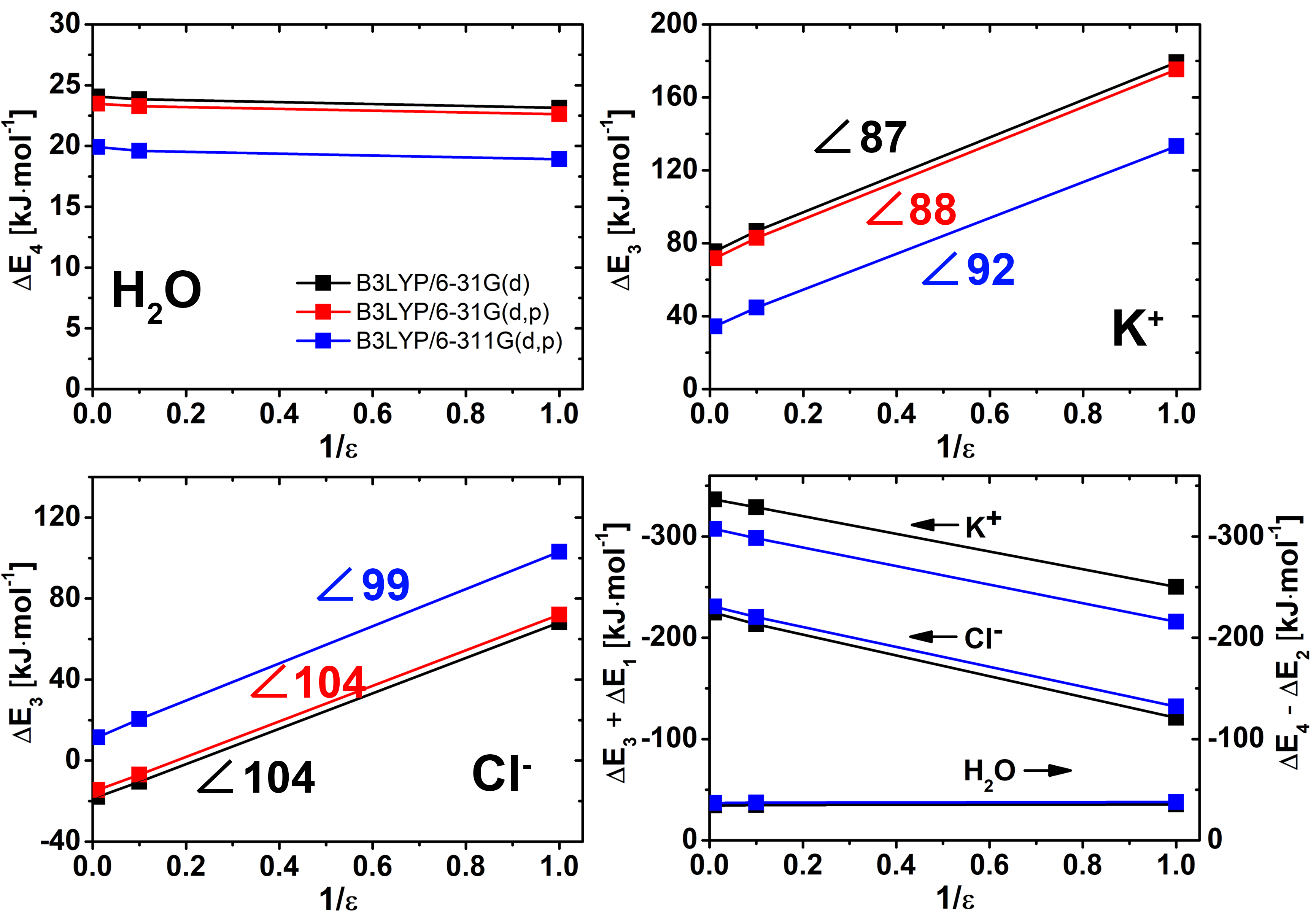}}
\caption{The effect of the basis set on the computed transfer energy of H\(_{2}\)O, K\(^{+}\) and Cl\(^{-}\) from  water bulk to water-filled CNT(5,5). The black curve is for the level of theory used in all structures in the paper, while red and blue lines represent improved levels of theory. The right bottom figure presents the same results as energies of transfer from vacuum phase to water filled CNT(5,5).}
\label{fig6}
\end{figure}
\textbf{Effect of the basis set on the dielectric energy.} Figure S6 displays the overall effect of the basis set used on the computed transfer energies. Addition of the p-function to hydrogen atoms has practically no effect on both the absolute values of the transfer energies and dependence on $\epsilon$. The use of triple-$\zeta$ basis set decreases the transfer energy for H\(_{2}\)O and K\(^{+}\) and increases it for Cl\(^{-}\). However, despite the difference of in transfer energies at 6-31G(d) and 6-311G(d,p) basis sets, the slopes of curves change only slightly. It appears the basis set mainly effects the computed absolute energies of binding to CNT, yet it insignificantly affects the contribution depending on $\epsilon$, i.e., the dielectric energy.
\newline

\textbf{The "surface energy" of water file termini in CNT(5,5).}  
The excess energy associated with two water file termini within CNT(5,5) may be estimated from the energies of 4- and 5-member files in CNT and of an empty CNT, as
\begin{equation} 
5E[(H_2O)_4 in CNT] - 4E[(H_2O)_5 in CNT] – E[CNT].
\end{equation}
In this expression the energy of transferring 20 water molecules to \textit{within} the file in a CNT and that of 5 CNTs is canceled out, leaving the terminal energy of a file in CNT only. The computed values of energies for all three structures entering this expression yield 34.6 kJ/mol for two termini or 17.3 kJ/mol per terminus, which is slightly above the energy of a missing hydrogen bond. 
\newline

\textbf{Chloride hydration in CNT (6,6).} Figure S7 shows the computed optimized structure of Cl\(^{-}\) surrounded by four water molecules with CNT(6,6). The result indicates that, unlike the analogous arrangement in CNT(5,5), chloride is no more residing in a single file and is hydrated by at least 3 water molecules.

\begin{figure}[h]
\center{\includegraphics[width=0.45\linewidth]{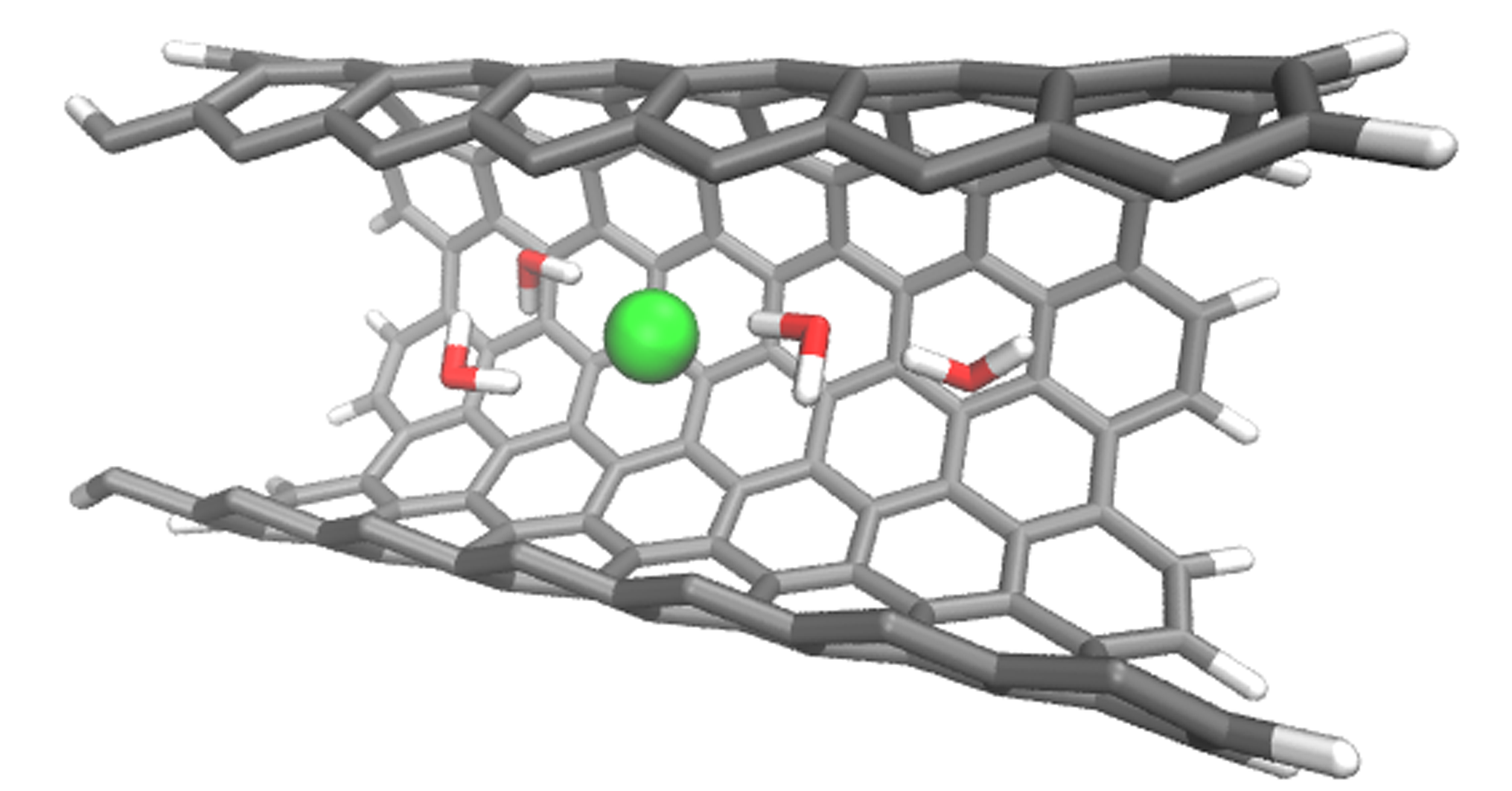}}
\caption{The optimized CNT(6,6) with diameter 0.8 nm and length 1.72 nm containing chloride ion surrounded by 4 water molecules at B3LYP/6-31G(d) level of theory in GAUSSIAN 09 Rev. B.01. in vacuum. Chloride ion, oxygen, hydrogen and carbon atoms are green, red, white and gray, respectively.}
\label{fig7}
\end{figure}